\documentclass[10pt]{article}
\usepackage{amsfonts}
\usepackage{epsfig}

\usepackage{amsmath}

\usepackage{color}

\begin{document}

\title{\LARGE \bf Brownian motion at the speed of light: a Lorentz
invariant family of processes}
\author{Maurizio Serva}
\maketitle
\centerline{\it Dipartimento di Ingegneria e Scienze dell'Informazione 
e Matematica, Universit\`a dell'Aquila, L'Aquila, Italy.}

\begin{abstract}

We recently introduced a new family of processes which describe particles
which only can move at the speed of light $c$ in the ordinary
3D physical space. The velocity, which randomly changes direction, 
can be represented as a point on the surface 
of a sphere of radius $c$ and its trajectories only may connect the 
points of this variety.
A process can be constructed both by considering jumps from one point to 
another (velocity changes discontinuously)
and by continuous velocity trajectories on the surface. 
We followed this second new strategy assuming that the velocity is described 
by a Wiener process (which is isotropic only in 
the 'rest frame')  on the surface of the sphere.
Using both Ito calculus and Lorentz boost rules, we succeed here in 
characterizing the entire Lorentz-invariant family of processes.
Moreover, we highlight and describe the short-term ballistic behavior 
versus the long-term diffusive behavior of the particles in the 3D physical 
space.

\end{abstract}

\begin{flushleft}
\end{flushleft}

\section{Introduction}

Brownian motion is a physical phenomenon which was historically 
modeled by the Wiener process, both directly identifying the particle
trajectories with the realizations of the process and indirectly assuming that
the particle velocity evolves according to a Ornstein-Uhlenbeck process. 
Although in the last century the Wiener process has been used
to describe a variety of phenomena in finance, biology, engineering, electronics 
and so on, its name remains strictly tied to the description of the motion 
of random particles.

If one tries to extend its use to the description of the random motion of 
relativistic particles, one clashes against one of its more characteristic
properties: trajectories are not differentiable, which means infinite speed
while relativity only allows luminal or subluminal velocities.
This fact doesn't imply that it is useless, on the contrary the
relativistic Brownian motion can be still modeled via a variety of modified Langevin 
equations which produce trajectories with a speed which is never superluminal.
We just quote  [1-19] 
which are a few of the studies which followed this strategy in the last fifty
years. 

In a recent research \cite{Serva:2020} we considered the extreme case in which 
a particle moves at  the speed of light, the aim was to produce a probabilistic 
tool which is related to the Brownian motion of light-speed particles in the 
same way as the Wiener process is related to the Brownian motion of classical
particles.
We do not derive here the erratic motion of a light-speed particle by some 
limit procedure which involves collisions with other particles or obstacles as in
\cite{Scalas:2015}, but we directly provide the mathematical framework.
Although it appears very difficult to imagine a physical device where a mass-less 
particle is trapped and scattered as a photon in a box of mirrors, this is, indeed,
closer to reality of what one could think.
Almost one century ago, Albert Einstein theoretically conceived a box in which a
single photon could be trapped in order to measure the relationship between mass 
and energy. 
Recently a team of physicists have created this box, a device that snares a photon
up to half a second \cite{Gleyzes:2007}.

One more reason for searching a Wiener description of relativistic random particles
is the possibility to extend the analogy between the Feynman integral
and the Wiener integral (Feynman-Kac formula) to the relativistic quantum domain.
The Schr\"odinger equation is solved by the Feynman integral
while the heat equation, which is connected to the first by analytic 
continuation, is solved by the Wiener integral. 
The relativistic versions of the Schr\"odinger equation are 
the Klein-Gordon (zero spin particles) and Dirac (spin one half) equations.
Both are hyperbolic equations (Dirac equation in its second order formulation).
Analytic continuation gives rise to elliptic equations,
the point is:  which process is associated to the elliptic equations?

The first answers were given in 
\cite{Ichinose:1982, Ichinose:1984, Jacobson:1984, Ichinose:1986, Ichinose:1987}
and later implemented in \cite{De Angelis:1990, De Angelis:1991} where
the Wiener process was still the main ingredient, but a four dimensional
one with both position and time following trajectories which are the realizations
of a Wiener process with the proper time as index. 
The proper time is then eliminated by a procedure based on hitting times.
Nevertheless, the resulting process is unphysical since the speed is 
not bounded and the whole construction only results in a tool for obtaining
a probabilistic solution of some elliptic equations.
If one forces the approach to the realm of physics one has to abandon 
Markov property and the single particle picture \cite{Serva:1988}.

There is a third way to approach the relativistic problem with
a process which is physical and allows to construct the solution
of the quantum hyperbolic equations.
In 1956 the Polish physicist and mathematician Mark Kac 
considered a (1+1)-dimensional process (one space dimension + time)
where the particle travels at speed of light
(left or right) and randomly inverts its velocity and he 
proved that the associated probability density
satisfies the telegrapher equation \cite{Kac:1956}.

About thirty years after the Kac pioneering work,
Gaveau {\it et al.} noticed that the telegrapher equation could be easily 
associated both to the Dirac equation in 1+1 dimensions (first order formulation)
and to the Klein-Gordon equation also in 1+1 dimensions 
(second order formulation).
Using this equivalence they were able to give a probabilistic solution
(by the backward Kolmogorov equation) to these fundamental
quantum equations \cite{Gaveau:1984}.
This result was later refined and extended in \cite{Blanchard:1986, Combe:1987}.
The weak point was that both Kac and these later constructions only
worked for particles in 1+1 dimensions .

Indeed, the process considered in \cite{Kac:1956,Gaveau:1984,Blanchard:1986,Combe:1987}
is part of a  larger class, in fact, by Lorentz boosts new processes can be obtained 
with particles moving at the speed of light  (a simple consequence of the fact that 
a light-speed particle in an inertial frame is also light-speed in any other inertial 
frame).
The processes of this larger class have in general an unbalanced 
probability rate of velocity inversion {\it i.e,}
the inversions from right to left occur with a different probability
rate of those from left to right, as a consequence, 
the particle may have a non vanishing average velocity.

The class of these one-dimensional light-speed processes was further extended 
by considering inversion rates which not only depend on the sign of the velocity
but also on position and time.
This extension gave the possibility to reformulate the quantum mechanics of a 
relativistic particle in terms of stochastic processes \cite{Serva:1986}
in the spirit of Nelson's stochastic mechanics \cite{Nelson:1967}.
Again, this construction was limited to 1+1 dimensions.

All the processes mentioned above can be related to the Feynman checkerboard,
which was also proposed as a mathematical tool for Fermions path integrals 
(see, for example, \cite{Kauffman:1996}).
The main difference is that in the (1+1)-dimensional Kac historical model and 
in the related ones (included the processes in this work) both time and space 
are continuous {\it ab initio}, while in the Feynman checkerboard they are 
both discrete.
In case,  continuity can be reached by a proper limit \cite{Molfetta:2012}.
Both types of approach have been used to define path integrals for Dirac equation,
but only the checkerboard can be placed at the origin of quantum walks
\cite{Molfetta:2012,Molfetta:2013,Jay:2018}.

In this paper we consider a family of processes which generalizes the Kac 
approach to the (3+1)-dimensional case (three space dimensions + time). 
The goal is to construct a Brownian motion
which is the most similar to the Wiener process among all those processes 
which do not conflict with relativity.  
We assume that the particle only moves at the speed of light $c$ which implies 
that velocity can be represented by a point on the surface of a sphere of 
radius $c$. 
We also assume that in the 'rest frame', the velocity performs a isotropic
Wiener process on that
surface (which corresponds to anisotropic Wiener processes in general frames).

In this way the speed is always $c$ which is the largest among those compatible 
with relativity, but velocity direction changes. 
It should be remarked 
the trajectories of the velocity are almost everywhere continuous 
but they are not differentiable, on the contrary the trajectories of the 
positions are continuous and differentiable. 

In the 'rest frame' the particle is ballistic at short times (position 
changes proportionally to time), and 
ordinarly diffusive ($E[x^2] \sim t$) at large times for which
the average velocity vanishes.
Then, one has to consider all the processes generated by Lorentz boosts.
The instantaneous velocity of these processes must be also luminal,
because a luminal particle is luminal in any inertial frame.
Therefore, the velocity still remains on the surface of the sphere,
nevertheless, its diffusion is anisotropic
and the average velocity is unvanishing at large times.
The construction of this family of processes, which transform
one into the other by Lorentz boost, needs Ito calculus
which leeds to the core equation (\ref{steq'}) which 
represents the entire family.

The paper is simply organized:
in section 2 we introduce the process in the 'rest frame'.
The velocity process on the sphere is formulated in a new and more
economic way which allow a simpler use of Ito calculus.
In section 3 we characterize the entire family of processes
generated by Lorentz boosts.
Nevertheless, the very long application of Ito calculus to reach this goal
is postponed in an Appendix that eventually the reader can skip.
Averages are computed in section 4 where we also 
highlight and describe the short-term ballistic behavior 
versus the long-term diffusive behavior of the particles in the 3D physical 
space.
Summary and  outlook can be finally found in section 5.

\section{Stochastic equations for the process in the 'rest frame'}

The particle velocity performs 
a isotropic Wiener process on the surface of a sphere of radius $c$,
In this way, while the velocity direction changes, the speed always 
equals $c$, which is the largest among those compatible with relativity. 

The equations governing this process (Ito notation) are:
\begin{equation}
\begin{aligned}
& d{\bf x}(t) = {\bf c}(t) dt , \\
& d{\bf c}(t) = -\omega^2 {\bf c}(t)\, dt + \omega c \, d {\bf w}(t)
\end{aligned}
\label{steq}
\end{equation}
where, according to Ito, $d{\bf c}(t) = {\bf c}(t+dt) - {\bf c}(t)$
and $d{\bf w}(t) ={\bf w}(t+dt)-{\bf w}(t)$
is a two component standard Wiener increment on the plane perpendicular to 
$\bf c$$(t)$ such that $E[\, |d{\bf w}(t)|^2]$ = $2dt$.

The second of the above equations, given that $d{\bf w}(t)$
is a two component Wiener increment tangent to the surface,
describes a isotropic Wiener process on that surface of a sphere.
This process was studied for the first time at least 70 years 
ago \cite{Yosida:1949,Yosida:1952}.

It is straightforward to verify that $|{\bf c} (t)|$ = $c$
at any time $t \ge 0$. In fact, according to Ito, one trivially obtains 
$d{\bf c}^2(t)=0$ (in next section this equality is explicitly proven for
the general family of processes generated by Lorentz boosts).
Therefore,
equation (\ref{steq}) describes a particle which has constant speed and whose 
velocity changes direction following continuous but not differentiable
trajectories. 

The present model can be seen as a 3 space dimensions version
of the Kac process. In fact, a constant speed process in one 
space dimension can be only constructed by considering jumps between the 
two possible velocities.
In three space dimensions, constant speed means that velocity is represented by 
a point on the surface of a sphere (with the speed as radius). A process over this 
surface can be realized by allowing jumps between points 
(velocity jumps from one value to another) or by continuous Wiener trajectories 
on the surface (velocity changes direction in a continuous way). 
There are only these two choices for a Markovian 
generalization of the Kac process to three space dimensions, 
we followed the second.
Nevertheless, the present model can be also considered as a peculiar 
Ornstein-Uhlenbeck process constructed in such a way that speed remains constant.

In the following pages we will omit the time as an explicit
argument when it is not strictly necessary.
For example, we will simply write ${\bf c}$, $d{\bf c}$, ${\bf w}$
and $d{\bf w}$ for ${\bf c}(t)$, $d{\bf c}(t)$, ${\bf w}(t)$
and $d{\bf w}(t)$.

The increment $d{\bf w}$ has to be a two 
component Wiener differential perpendicular to $\bf c$
(which means tangent to the surface), 
nevertheless, the  recipe for its construction is not univocal. 

In the early seventies Strook and then Ito \cite{Stroock:1971,Ito:1975} 
constructed the increment $d\bf  w$ in the second of the equations (\ref{steq}) 
as
\begin{equation}
d{\bf w} =
\left(\mathbb{ I}- \mathbf{n}\mathbf{n}^\mathsf{T} \right) \,d{\bf W}
\label{sigma}
\end{equation}
where ${\bf n}(t)= {\bf c}(t)/{c}$ is a time dependent unitary vector,
${\bf W}$ is a standard three dimensional Wiener process, 
$\mathbb{ I} $ is the $3 \times 3$ identity matrix and the row vector
$\mathbf{n}^\mathsf{T}$ is the transposed of the column vector $\bf n$.
One gets $d{\bf c} = -\omega^2  {\, \bf c}\, dt +  \sigma \,d{\bf W}$
where
$\sigma= 
\omega c \left( \mathbb{I}-\mathbf{n}\mathbf{n}^\mathsf{T} \right) $
is a $3 \times 3$ matrix.

About ten years later a simpler choice was considered \cite{Price:1983,Berg:1985}:
\begin{equation}
d{\bf w} =
{\, \bf n} \times  \,d{\bf W},
\label{sigma'}
\end{equation}
which leads to $ d{\bf c} =-\omega^2  {\, \bf c}\, dt + \hat \sigma \,d{\bf W} $
where 
$\hat \sigma= \omega c [{\bf n}]$ with $[{\bf n}]$ being
the $3 \times 3$ skew matrix representation of the vector $\bf n$.
It is easy to check that 
$\sigma \, \sigma^\mathsf{T} =  \hat \sigma \, {\hat \sigma}^{\mathsf{T}}
= \omega^2 c^2 \left(\mathbb{ I}-\mathbf{n}\mathbf{n}^\mathsf{T}\right) $ 
which implies that 
the Forward Kolmogorov Equation is the same for choices (\ref{sigma})
and (\ref{sigma'}).
See  \cite{Brillinger:1997,Krishna:2000} for properties and applications.

Both implementations of the two-dimensional increment $d{\bf w}$ are made by a 
three dimensional Wiener process $\bf W$$(t)$, which is somehow
redundant for the construction of a two-dimensional increment.

We propose here to use in place of the standard three-components Wiener process 
$\bf W$ = $(W_1, \, W_2, \, W_3)$, a standard two-components
Wiener processes $w_2, w_3$ (we write $w_2,\,  w_3$ in place of
$W_2, \, W_3$ to avoid confusion).
Our choice is 
\begin{equation}
d{\bf w}= 
{\bf n_2}\, d w_2 + {\bf n_3}\, dw_3
\label{m}
\end{equation}
where ${\bf n_2}(t)$ and ${\bf n_3}(t)$ are two unitary vectors
perpendicular each other and also perpendicular to  ${\bf n}(t)$. 
The Wiener increments are independent which
implies $E[dw_1(t) \, dw_2(t)] = 0$ and they are standard
which means $E[(d{\bf w}(t))^2]$ = $2dt$. 

It must be clear that the orientation of ${\bf n_2}(t)$ and ${\bf n_3}(t)$ 
can be arbitrarily chosen on the plane perpendicular to ${\bf n}(t)$.
In the following we make a choice which is motivated by the fact that
it is the simplest for our goal, which is to construct, by Ito calculus,
the general family of processes generated by Lorentz boosts.
In \cite{Serva:2020} we made a totally different choice which allowed us
to separate the spacial variables from the velocity variables
in order to write down a Forward Kolmogorov Equation 
directely in a 3D configuration space.

Given a constant vector ${\bf v}$,  we chose here
\begin{equation}
{\bf n_2}= \frac{{\bf v} \times {\bf n}}{| {\bf v} \times {\bf n}|},
\,\,\,\,\,\,\,\,\,\,\,\,\,\,\,\,\,\,\,
{\bf n_3}= {\bf n} \times {\bf n_2},
\label{nn}
\end{equation}
so that $\bf v$, $\bf n$=$\bf c$/c and $\bf n_3$  are on the same plane
and $\bf n_2$ is perpendicular to it.
To fix the ideas one can put the north pole in the $\bf v$ direction
with respect to the center of the sphere,
so that ${\bf n_3}$ is tangent to a meridian and points to north,
while ${\bf n_2}$ is tangent to a parallel and points to est.
In this way it is simple to pass to spherical coordinates.
At the poles (where $\bf n$  equals $\pm$ $ \bf v$/$| \bf v|$)  
the unitary vectors  ${\bf n_2}$ and ${\bf n_3}$ 
can be arbitrarily chosen perpendicularly to $\bf v$.

Equations (\ref{sigma}), (\ref{sigma'}) and (\ref{m})
are three totally equivalent way to 
construct the same  Wiener increment tangent to the surface.
According to our representation (\ref{m}), the second equation in (\ref{steq}) 
rewrites as:
\begin{equation}
d{\bf c} = -\omega^2 {\bf c}\, dt + \omega c \, 
({\bf n_2}\, d w_2 + {\bf n_3}\, dw_3),
\label{steq2}
\end{equation}
the advantage being that we use only a two component Wiener process in
place of a three component one, moreover
this stochastic equation is straightforwardly associated
to the velocity spherical Laplacian in the Kolmogorov Equations
when it is expressed in terms of longitude and latitude.

We stress again that our specific choice (\ref{nn}) is only dictated by 
convenience for later calculations in this paper, any other choice which 
keeps $\bf n_2$ and $\bf n_3$ perpendicular to $\bf n$ and perpendicular 
each other is equally admissible.

\section{Lorentz boosts and stochastic equations in a generic inertial frame}

In the 'rest frame' the  velocity $\bf c$$(t)$ of the 
particle evolves according to equation (\ref{steq2}) where ${\bf n_2}$ 
and ${\bf n_3}$ are defined by (\ref{nn}). 
Then, assume that this 'rest frame' moves at constant velocity $\bf u$ 
(without rotating) with respect  to a second inertial frame. 
Since the choice of $\bf v$ in (\ref{nn}) is arbitrary, we can leave it
to coincide with  $\bf u$. In the next we will only use $\bf v$
to indicate both the velocity in (\ref{nn}) and the velocity of the 'rest frame'.

According to special relativity, the velocity $\bf c'$$(t)$ of the particle in 
the second frame is
\begin{equation}
{\bf c'}(t)= \frac{1}{1+\frac{{\bf v}\cdot{\bf c}(t)}{c^2}}
\left[\alpha{\bf c}(t)+{\bf v} + (1-\alpha)\frac{{\bf v} 
\cdot {\bf c}(t)}{v^2}{\bf v} \right]
\label{c'}
\end{equation}
where $\bf v$, $c$ and 
$\alpha =\left(1-\frac{v^2}{c^2} \right)^{\frac{1}{2}}$ are constant.
Notice that in the above equation the argument of ${\bf c'}$ is still $t$
and not the time $t'$ of the second frame.

By special relativity the velocity in this second inertial frame is also 
luminal (${\bf |c'|}=c$) (at the end of this section we will show 
that indeed $({\bf c'})^2= {\bf c}^2 =c^2$). 
If one also takes into account that the time increment $dt'$ in
the second frame  satisfies
\begin{equation}
\frac{dt}{dt'}= \frac{1}{\alpha}  
\left( 1-\frac{{\bf v} \cdot {\bf c'}}{c^2} \right),
\label{tt'} 
\end{equation}
one should be able to write from (\ref{c'})
and (\ref{tt'}) a stochastic equation for ${\bf c'} (t')$
analogous to (\ref{steq2}) and (\ref{nn}). 
Notice that in the new equation the time will be $t'$ and the increment 
 $d{\bf c'} = {\bf c'}(t'+dt') - {\bf c'}(t')$ must be expressed 
in terms of  ${\bf c'}(t')$,  $dt'$ and of the increments
$dw'_2 = w_2(t'+dt') - w_2(t')$ and $dw'_3 = w_3(t'+dt') - w_3(t')$.
In the second frame the particle will still
instantaneously move at the speed of light but, contrarily to the case of the process
in the 'rest frame', its  average velocity will not vanish at large times.

We define $\delta {\bf c'} = {\bf c'}(t+dt) - {\bf c'}(t)$
(notice the difference with $d{\bf c'} = {\bf c'}(t'+dt') - {\bf c'}(t')$), 
then a long and tedious application of Ito calculus (see the Appendix) leads to
\begin{equation}
\delta {\bf c'}=  - \frac{\omega^2}{\alpha^2} 
\left[ 1-\frac{{\bf v} \cdot {\bf c'}}{c^2} \right]^2 {\bf c'}  dt
+ \frac{\omega \, c}{\alpha}  
\left( 1-\frac{{\bf v} \cdot {\bf c'}}{c^2} \right) 
({\bf n'_2}\, d w_2 + {\bf n'_3}\, dw_3)
\label{ito}
\end{equation}
where $\bf n'_2$ and $\bf n'_3$ are the two unitary vectors perpendicular
to $\bf c'$ defined as in (\ref{nn}) (with $\bf n$, $\bf n_2$ and $\bf n_3$ 
replaced by $\bf n'$ = $\bf c'$$/c$, $\bf n'_2$ and $\bf n'_3$).
Then, taking into account (\ref{tt'})
and remembering that $dw_3 /dw'_3 = dw_2/dw'_2 =(dt/dt')^\frac{1}{2}$ one gets

\begin{equation}
d{\bf c'}=  - \frac{\omega^2}{\alpha^3} 
\left[ 1-\frac{{\bf v} \cdot {\bf c'}}{c^2} \right]^3 {\bf c'}  dt' 
+ \omega \, c\,  \left[ \frac{1}{\alpha}  
\left( 1-\frac{{\bf v} \cdot {\bf c'}}{c^2} \right) 
\right]^{\frac{3}{2}}  d {\bf w'}
\label{steq'}
\end{equation}
where $d {\bf w'}= {\bf n'_2}\, d w'_2 + {\bf n'_3}\, dw'_3$ is a two component 
increment perpendicular to ${\bf c'}$ in complete analogy with the 
process in the 'rest frame'. Notice that by (\ref{c'}) the
three vectors $\bf v$, $\bf c$ and $\bf c'$ are on the same plane
so that  $\bf n_3$, $\bf n'_3$ also are 
on the same plane. As a consequence $\bf n_2$ and $\bf n'_2$ are both
perpendicular to that plane  so that $\bf n_2$ = $\bf n'_2$.

This is the core equation, each process of the family is labeled by 
the index ${\bf v}$, the case ${\bf v}=0$ corresponds to the
process in the 'rest frame'.

One can easily prove from (\ref{steq2}) that the speed of the
particle remains constantly luminal {\it i.e,} $ d|{\bf c'}(t)| =0$, 
in fact by Ito calculus
\begin{equation}
d|{\bf c'}|^2 = - 2 \, \frac{\omega^2}{\alpha^3} 
\left[ 1-\frac{{\bf v} \cdot {\bf c'}}{c^2} \right]^3 |{\bf c'}|^2 dt' 
+ 2 \, c^2 \,\frac{\omega^2}{\alpha^3} 
\left[ 1-\frac{{\bf v} \cdot {\bf c'}}{c^2} \right]^3  dt'
+2 \frac{c \, \omega^3}{\alpha^3} \, 
\left[ 1-\frac{{\bf v} \cdot {\bf c'}}{c^2} \right]^3
{\bf c'} \cdot d {\bf w'}
\end{equation}
where the second term at the right comes from the second order 
contribution to Ito differential. 
Since ${\bf c'}$ and $d {\bf w'}$ are perpendicular the equation  
reduces to
\begin{equation}
d|{\bf c'}|^2 = - 2 \, \frac{\omega^2}{\alpha^3} 
\left[ 1-\frac{{\bf v} \cdot {\bf c'}}{c^2} \right]^3 
\left(|{\bf c'}|^2 - c^2 \right) dt' =0
\end{equation}
where the last equality holds if the initial velocity is luminal
{\it i.e,} $|{\bf c}(0)|=c$.

The fact that the process remains luminal is not
astonishing since a particle moving at the speed of light
also moves at the speed of light in any other inertial frame.
Therefore, the equation (\ref{steq'}) defines a family of 
light-speed processes (indexed by ${\bf v}$)
which transform one in the other by Lorentz boost.

We already mentioned that in a generic inertial frame (indexed by ${\bf v}$) 
the particle has 
a long term unvanishing average velocity. This is simple consequence 
of the fact that the rest frame moves at velocity ${\bf v}$ with respect to the 
generic frame and of the fact that the long term average velocity vanishes 
for the process in the 'rest frame' (we will prove this in the next section).

The reason of this long term average lies in the fact that 
the diffusion of ${\bf c'}$
slows down the more  ${\bf v} \cdot {\bf c'}$ is  large
as it can be inferred from equation (\ref{steq'}), 
This means that the particle spends more time with values of 
${\bf c'}$ aligned with ${\bf v}$ and less time when it is 
anti-aligned.
This is exactly the same one has with the (1+1)-dimensional
Kac process associated to the telegrapher equation.
In that simpler case a non vanishing average velocity
is determined by the fact that unbalanced rates of inversions
lead to a longer permanence of the velocity in one of the two directions.

\section{Short-term ballistic behavior versus long-term diffusive behavior}

In this section we only consider the process in the 'rest frame',
all results concerning averages 
can be eventually Lorentz transformed for the processes in a generic
inertial frame.

The stochastic equation (\ref{steq}) can be recast in an integral 
equation:
\begin{equation}
\begin{aligned}
& {\bf x}(t) = {\bf x}(0) +\int_0^t {\bf c}(s) ds, \\
& {\bf c}(t) = e^{-\omega^2 t} \left[  {\bf c}(0) +
\omega \, c \int_0^t e^{\omega^2 s} d {\bf w}(s) \right].
\end{aligned}
\label{sol}
\end{equation}
This is not a solution because $d {\bf w}(s)$, according to (\ref{m})
and (\ref{nn}), depends on ${\bf c}(t)$. 	
Let us mention that the proof that the particle velocity remains constantly 
luminal {\it i.e,} $ |{\bf c}(t)| =c$ can be eventually also obtained by the 
second integral equation in (\ref{sol}).

Starting from the second integral equation in
(\ref{sol}) one can easily find out that 
the following averages hold for $t \ge s \ge 0 $:
\begin{equation}
\begin{aligned}
& E[{\bf c}(t)] = e^{-\omega^2 t} {\bf c}(0), \\
& E[{\bf c}(t) \cdot {\bf  c}(s)] = c^2 \, e^{-\omega^2 (t+s)} 
+ 2\omega^2 c^2 e^{-\omega^2 (t+s)}\int_0^s e^{2\omega^2 u} du
=c^2 \, e^{-\omega^2 (t-s)}.
\end{aligned}
\end{equation}
Notices that the first of the equalities above says that the average velocity 
$E[{\bf c}(t)]$ vanishes for large $t$ however, for a generic inertial 
frame, $E[{\bf c'}(t')]$ doesn't vanish for large $t'$.
This is simple consequence 
of the fact that the rest frame moves at velocity ${\bf v}$ with respect to the 
generic frame.

Using these averages and the first of the integral equations in
(\ref{sol}), one also obtains 
\begin{equation}
\begin{aligned}
& E[{\bf x}(t)] 
= {\bf x}(0)+ \frac{1 \!- \! e^{-\omega^2 t}}{\omega^2} \, {\bf c}(0), \\
& E[ \left({\bf x}(t)-{\bf x}(0)\right)^2 ] = 
2 c^2 \int_0^t \int_0^s e^{-\omega^2 (s-u)} du \, ds=
\frac{2 c^2}{\omega^2} \, t -
\frac{2c^2}{\omega^4} \left( 1\!- \! e^{-\omega^2 t}  \right).
\end{aligned}
\label{xx}
\end{equation} 
The above averages imply, for large times, a diffusive behavior with 
coefficient $\frac{ c^2}{\omega^2}$, in this limit
one has in fact 
\begin{equation}
E[ \left({\bf x}(t)-{\bf x}(0)\right)^2 ]
 \sim \frac{2 c^2}{\omega^2} \, t,
\label{long}
\end{equation} 
on the contrary,  for short times one has
\begin{equation}
E[{\bf x}(t)-{\bf x}(0)] \sim {\bf c}(0) t, \,\,\,\,\,\,\,\,\,\,\,\,\,\,\,
E[ \left({\bf x}(t)-{\bf x}(0)\right)^2 ] \sim c^2 t^2,
\label{short}
\end{equation} 
which means ballistic behavior at the speed of light.
 
The short-term ballistic behavior it is not completely unexpected. In fact,
for a small time $\Delta t \ll 1/\omega^2$ the velocity of a particle remains 
almost constant. This can be understood from the second equation in
(\ref{steq}) which, having defined 
$\Delta{\bf c} ={\bf c}(\Delta t)-{\bf c}(0)$, implies $|\Delta{\bf c}| 
\approx|\omega^2 {\bf c}\,\Delta t - \omega c \, {\bf w}(\Delta t)|
\le \omega^2 c \Delta t + \omega c |{\bf w}(\Delta t)|$.
Since $|{\bf w}(\Delta t)|$ is of the order of $ \sqrt{\Delta t} $ and given 
that $\Delta t \ll 1/\omega^2$,
one finally obtains $|\Delta{\bf c}| /c \ll 1$. 
This short-term ballistic 
behavior was already observed and described in depth in \cite{Debbasch:2012}
for the Relativistic Ornstein-Uhlenbeck Process \cite{Debbasch:1997}. 
In case, it could be also coarsely derived from equation (2) in 
\cite{Debbasch:2012} by adapting the line of reasoning we followed above.

\section{Summary and outlook}

In conclusion we have found that equation (\ref{steq'}) describes a family of 
light-speed processes which transform one in the other by Lorentz boost.
Their main characteristics can be resumed as follows:

\begin{itemize}

\item
the family of processes that we propose generalizes to 3+1 dimensions
the 1956 idea of Mark Kac in the sense that particles only move at 
the speed of light.
Although the Kac process can be generalized to 3+1 dimensions in a different 
way, for example considering a velocity which performs jumps in place of having 
continuous trajectories, we think the process presented here,
having as a constitutive ingredient the Wiener process, is the most natural choice 
for this generalization. Moreover, since the speed is always the maximum possible
given the relativistic constraint,
it posseses the trajectories which better mimics the (infinite speed) Wiener 
trajectories;

\item
the long term average velocity vanishes in the 'rest frame',
but it does not in a generic frame.
This is is a consequence of the fact that
according to  (\ref{steq'}) the diffusion of the velocity 
slows down the more  ${\bf v} \cdot {\bf c'}$ is  large.
In turn, this means that the particle spends more time with values of 
${\bf c'}$ aligned with ${\bf v}$ and less time when it is 
anti-aligned.
This is exactly the same situation  one has with the (1+1)-dimensional
Kac process associated to the telegrapher equation since the probability rate
of inversion of velocity can be different for left/right  and right/left 
inversions \cite{Serva:1986};

\item
for large  times the behavior of the position is diffusive with coefficient 
$\frac{ c^2}{\omega^2}$,  one has in fact $E[ \left({\bf x}(t)-{\bf x}(0)\right)^2]
 \sim \frac{2 c^2}{\omega^2} \, t$. On the contrary,  for short times
$E[{\bf x}(t)-{\bf x}(0)] \sim c t$ and
$E[ \left({\bf x}(t)-{\bf x}(0)\right)^2 ] \sim c^2 t^2$
which means ballistic behavior at the speed of light.
The short-term ballistic behavior holds for $t$ smaller than $1/\omega^2$.

\end{itemize}

This process is the natural candidate for modeling the Brownian motion 
of mass-less particles, nevertheless, its use should not be limited to this case.
The situation is similar in the non-relativistic realm; thought
a particle with infinite speed is unphysical, the Wiener process
is largely used to model its erratic movement. 
Another point that deserves investigation and which contributed to prompt
this work is the possible connection of its Backward Kolmgorov Equation 
with relativistic equations as Klein-Gordon and Dirac. 
The goal would be to find a generalization of the Gaveau {\it et al.} 
approach to the (3+1)-dimensional case. This topic is presently under study.

\section*{Appendix: Ito calculus}

In this appendix we apply Ito calculus, in order to obtain equation (\ref{ito})
from equation (\ref{c'}). 
Since we defined $\delta {\bf c'} = {\bf c'}(t+dt) - {\bf c'}(t)$, 
then from  (\ref{c'}) we immediately get
\begin{equation}
\delta{\bf c'}= 
\frac{1}{1+\frac{{\bf v}\cdot({\bf c}+d{\bf c})}{c^2}}
\left[\alpha({\bf c}+d{\bf c})+{\bf v} + (1-\alpha)\frac{{\bf v}
\cdot ({\bf c}+d{\bf c})}{v^2}{\bf v} \right]-
\frac{1}{1+\frac{{\bf v}\cdot{\bf c}}{c^2}}
\left[\alpha{\bf c}+{\bf v} + (1-\alpha)\frac{{\bf v}
\cdot {\bf c}}{v^2}{\bf v} \right]
\end{equation}
where $ {\bf c}  +  d{\bf c}  = {\bf c}(t+dt) $ and ${\bf c} = {\bf c}(t)$.
This is still not a Ito increment, but it is the trivial
application of the definition $\delta {\bf c'} = {\bf c'}(t+dt) - {\bf c'}(t)$.
This equation can be exactly rewritten as
\begin{equation}
\delta{\bf c'}=  
\frac{d \epsilon}{1+d \epsilon} \left[ (1-\alpha)\frac{c^2}{v^2}{\bf v}-
\frac{1}{1+\frac{{\bf v}\cdot{\bf c}}{c^2}}
\left(\alpha{\bf c}+{\bf v} + (1-\alpha)\frac{{\bf v}
\cdot {\bf c}}{v^2}{\bf v} \right)\right]  +
\frac{1}{1+d \epsilon} \frac{\alpha{d\bf c}}{1+\frac{{\bf v}\cdot{\bf c}}{c^2}}
\label{dc'}
\end{equation}
where $d{\bf c}$ is given by (\ref{steq2}) and where
\begin{equation}
d \epsilon=\frac{1}{1+\frac{{\bf v}\cdot{\bf c}}{c2}} \,
\frac{ {\bf v}\cdot d{\bf c}}{c^2} = -\frac{\omega^2 }{c^2} 
\frac{ {\bf v}\cdot {\bf c}}{1+\frac{{\bf v}\cdot{\bf c}}{c2}} 
\, dt +  \frac{\omega }{c} \,
\frac{ {\bf v}\cdot {\bf n_3}}{1+\frac{{\bf v}\cdot{\bf c}}{c2}} \, dw_3.
\label{eps}
\end{equation}

The next step is to calculate the right side of the equation (\ref{dc'})
in terms of the new 
variables ${\bf c'}$, ${\bf n_2'}$ and ${\bf n_3'}$. First of all, 
using (\ref{c'}) we immediately rewrite the equation (\ref{dc'}) as
\begin{equation}
\delta {\bf c'} = \frac{1}{1+d \epsilon} 
\left[d \epsilon \,  \left( (1-\alpha)\frac{c^2}{v^2}{\bf v} - {\bf c'} \right)
+  \frac{ 1-\frac{{\bf v}\cdot{\bf c'}}{c^2} }{\alpha}\, d{\bf c}\, \right],
\label{2dc'}
\end{equation}
but we also need to calculate $d \epsilon$ and $ d{\bf c}$ in 
terms of the new coordinates. In order to reach this goal we need to recall that
\begin{equation}
{\bf c}= \frac{1}{1-\frac{{\bf v}\cdot{\bf c'}}{c^2}}
\left[\alpha{\bf c'}-{\bf v} + (1-\alpha)\frac{{\bf v}
\cdot {\bf c'}}{v^2}{\bf v} \right] ,
\,\,\,\,\,\,\, \to \,\,\,\,\,\,\,
{\bf v} \cdot {\bf c}= \frac{1}{1-\frac{{\bf v}\cdot{\bf c'}}{c^2}}
\left[ {\bf v}\cdot {\bf c'} - v^2 \right] \, ,
\label{c}
\end{equation}
moreover
\begin{equation}
c \, {\bf n_3}=  \frac{1}{1-\frac{{\bf v}\cdot{\bf c'}}{c^2}}
\left[\alpha c {\bf n_3'}-{\bf v} \times {\bf n_2'} + 
(1-\alpha)\frac{{\bf v}
\cdot {\bf c'}}{v^2} \, {\bf v} \times {\bf n_2'}\right],
\,\,\,\,\,\,\, \to \,\,\,\,\,\,\, 
{\bf v} \cdot {\bf n_3}= \alpha \frac{{\bf v} 
\cdot {\bf n_3'}}{1-\frac{{\bf v}\cdot{\bf c'}}{c^2}}.
\label{n3}
\end{equation}
From these two last equations one easily realize
that the three vectors $\bf c$, $\bf v$ and  $\bf c'$ are co-planar.
As well, ${\bf n_3}$ and ${\bf n_3'}$ lie on the same plane. 
Moreover, ${\bf n_2'}={\bf n_2}$ is perpendicular to that plane.
We get

\begin{equation}
d \epsilon = -\frac{\omega^2 }{c^2} 
\frac{ {\bf v}\cdot {\bf c'}-v^2}{\alpha^2} 
\, dt +  \frac{\omega }{c} \,
\frac{ {\bf v}\cdot {\bf n_3}'}{\alpha} \, dw_3,
\label{eps'}
\end{equation}
where $d {\bf c}$ is given by (\ref{steq2}) with
$\bf n_2$=$\bf n_2'$ and $\bf c$ and $\bf n_3$ given respectively by 
the first equation in (\ref{c}) and the first equation in (\ref{n3}).

We are now ready compute the Ito differential {\it i.e,} we are ready
to rewrite the differential $\delta {\bf c'}$ keeping only terms of order $dt$.
To obtain this result we have first of all to expand (\ref{2dc'}) 
to the second order with respect to the differentials:

\begin{equation}
\delta {\bf c'} \approx
\left[d \epsilon \,  \left( (1-\alpha)\frac{c^2}{v^2}{\bf v} - {\bf c'} \right)
+  \frac{ 1-\frac{{\bf v}\cdot{\bf c'}}{c^2} }{\alpha}\, d{\bf c}\,\, \right] - 
\left[(d \epsilon)^2 \,  \left( (1-\alpha)\frac{c^2}{v^2}{\bf v} - 
{\bf c'} \right) +  d \epsilon d{\bf c}] \, 
\frac{ 1-\frac{{\bf v}\cdot{\bf c'}}{c^2} }{\alpha}\, \right],
\label{3dc'}
\end{equation}
then, we have replace the second order differentials
$(d \epsilon)^2$ and $d \epsilon \, d{\bf c}$
by the terms proportional to $dt$ of their averages:
\begin{equation}
(d \epsilon)^2 \approx
\left(\frac{\omega }{c} \right)^2 \,
\frac{({\bf v} \cdot {\bf n_3'})^2}{\alpha^2}\, dt \, ,
\,\,\,\,\,\,\,\,\,\,\,\,\,\,\,\,\,
d \epsilon d{\bf c} \approx
\frac{\omega^2 }{\alpha}  \,
({\bf v} \cdot {\bf n_3'}) {\bf n_3}\, dt
\end{equation}
where $\bf n_3$ must be expressed in terms of the new variables
by the first equation in (\ref{n3}).
 
Let us rewrite equation (\ref{3dc'}) as
\begin{equation}
\delta {\bf c'}= \delta \mathcal{A}+ \delta \mathcal{B}+ \delta \mathcal{C}
\label{4dc'}
\end{equation}
where $\delta \mathcal{A}$ is the term proportional to $dt$ which comes from
the first order differentials (the deterministic part of the first
term between square parenthesis in (\ref{3dc'})), 
$\delta \mathcal{B}$ is the term proportional to $dt$ which 
comes from the second order differential (the second
term between square parenthesis in (\ref{3dc'})) and
$\delta \mathcal{C}$ is the random term (the random part of the first
term between square parenthesis in (\ref{3dc'})).
After having expressed all the old variables in terms of the new ones
(except $\bf n_3$), we have
\begin{equation}
\begin{aligned}
& \delta \mathcal{A}=   -\left[\frac{\omega^2 }{c^2} 
\frac{ {\bf v}\cdot {\bf c'}-v^2}{\alpha^2}
\left( (1-\alpha)\frac{c^2}{v^2}{\bf v} - {\bf c'} \right)
+  \frac{\omega^2}{\alpha}\, 
\left(\alpha{\bf c'}-{\bf v} + (1-\alpha)\frac{{\bf v}
\cdot {\bf c'}}{v^2}{\bf v} \right) \,\, \right]dt , \\ 
& \delta \mathcal{B}=    -\frac{\omega ^2}{\alpha^2} ({\bf v} \cdot {\bf n_3'})
\left[\frac{{\bf v} \cdot {\bf n_3'}}{c^2}\, \,  
\left( (1-\alpha)\frac{c^2}{v^2}{\bf v} - 
{\bf c'} \right) +  
\left( 1-\frac{{\bf v}\cdot{\bf c'}}{c^2}\right)\,  {\bf n_3}\right] dt ,    \\
& \delta \mathcal{C}= \left[ \frac{\omega }{c} \,
\frac{ {\bf v}\cdot {\bf n_3}'}{\alpha}  \,  
\left( (1-\alpha)\frac{c^2}{v^2}{\bf v} - {\bf c'} \right)\, dw_3
+  \frac{ 1-\frac{{\bf v}\cdot{\bf c'}}{c^2} }{\alpha}\, 
\omega c \left( {\bf n'_2} \, dw_2 + {\bf n_3} \,dw_3\right) \right] 
\end{aligned}
\label{5dc'}
\end{equation}
with $\bf n_3$ given by (\ref{n3}) in terms of the new variables.
After some rearrangement of terms we get:
\begin{equation}
\delta \mathcal{A} = 
-\left[\frac{\omega^2 }{c^2} 
\frac{ {\bf v}\cdot {\bf c'}}{\alpha^2}
- \frac{\omega^2}{\alpha^2}\, \right]{\bf v} dt
+\left[\frac{\omega^2 }{c^2} 
\frac{ {\bf v}\cdot {\bf c'}-c^2}{\alpha^2}
\right]{\bf c'}dt
=\frac{\omega^2 }{\alpha^2} \left[ 1-\frac{{\bf v}\cdot{\bf c'}}{c^2}
\right]({\bf v}-{\bf c'})dt.
\end{equation}
This differential lies in the plane of $\bf c'$ and $\bf n_3'$
and can be decomposed along these two vectors:

\begin{equation}
\delta \mathcal{A} 
=-\frac{\omega^2 }{\alpha^2} \left[ 1-\frac{{\bf v}\cdot{\bf c'}}{c^2}
\right]^2 {\bf c'}dt+\frac{\omega^2 }{\alpha^2} 
\left[ 1-\frac{{\bf v}\cdot{\bf c'}}{c^2}
\right]({\bf v} \cdot{\bf n_3'}){\bf n_3'}dt.
\end{equation}
Analogously, the term $\delta \mathcal{B}$, after decomposition along
$\bf c'$ and $\bf n_3'$, can be rewritten as

\begin{equation}
\begin{aligned}
\delta \mathcal{B} = -\frac{\omega ^2}{\alpha^2} ({\bf v} \cdot {\bf n_3'}) &
\left( 1-\frac{{\bf v}\cdot{\bf c'}}{c^2}\right)\,
\left[\frac{{\bf v} \cdot {\bf n_3'}}
{ 1-\frac{{\bf v}\cdot{\bf c'}}{c^2}}\, \,  
\left( (1-\alpha)\frac{{\bf v} \cdot {\bf c'}}{v^2} - 1 \right) +  
  {\bf n_3} \cdot {\bf c'}\right] \frac{{\bf c'}}{c^2} dt \\
 & -\frac{\omega ^2}{\alpha^2} ({\bf v} \cdot {\bf n_3'})
\left( 1-\frac{{\bf v}\cdot{\bf c'}}{c^2}\right)\,
\left[\frac{{\bf v} \cdot {\bf n_3'}}
{\left( 1-\frac{{\bf v}\cdot{\bf c'}}{c^2}\right)\,}\, \,  
\left( (1-\alpha)\frac{{\bf v} \cdot {\bf n_3'}}{v^2}  \right) +  
  {\bf n_3} \cdot {\bf n_3'}\right] {\bf n_3'} dt .
\end{aligned}
\label{dd}
\end{equation}
Since ${\bf c'} \cdot {\bf n_3}$=$-{\bf c} \cdot {\bf n_3'}$, from (\ref{c}) 
by scalar product with ${\bf n_3'}$, one gets 

\begin{equation}
{\bf c'} \cdot {\bf n_3}=  \frac{1}{ 1-\frac{{\bf v}\cdot{\bf c'}}{c^2}}
\left[1 - (1-\alpha)\frac{{\bf v} \cdot {\bf c'}}{v^2 c^2}\right]
{\bf v} \cdot {\bf n_3'} ,
\label{vw}
\end{equation}
therefore the first term at the right side
of equality (\ref{dd}) vanishes, moreover from (\ref{c}) 
and by the definitions of $\bf n_3$ and $\bf n_3'$ one has that

\begin{equation}
{\bf n_3} \cdot {\bf n_3'}= \frac{{\bf c} \cdot {\bf c}'}{c^2}= 
\frac{1}{1-\frac{{\bf v}\cdot{\bf c'}}{c^2}}
\left[\alpha - \frac{{\bf v}\cdot{\bf c'}}{ c^2} + 
(1-\alpha)\frac{({\bf v}\cdot {\bf c'})^2}{v^2}\right]
\label{cc'} ,
\end{equation}
which, can be substituted in the second term of (\ref{dd}) in order
to obtain

\begin{equation}
\delta \mathcal{B} = 
-\frac{\omega ^2}{\alpha^2} ({\bf v} \cdot {\bf n_3'})
\left( 1-\frac{{\bf v}\cdot{\bf c'}}{c^2}\right)\,{\bf n_3'} dt .
\end{equation}
A scalar product of the third of (\ref{5dc'}) with  ${\bf c'}$ gives
\begin{equation}
{\bf c'} \cdot \delta \mathcal{C}=
\frac{\omega c}{\alpha} \left[ \,
{\bf v}\cdot {\bf n_3}' \,  
\left( (1-\alpha)\frac{{\bf v} \cdot{\bf c'}}{v^2} - 1\right)\,
+  \left( 1-\frac{{\bf v}\cdot{\bf c'}}{c^2} \right)\, 
\left( {\bf c'} \cdot {\bf n_3} \right) \right]  dw_3 =0
\end{equation}
where the  equality is obtained using (\ref{vw}).
Construction is coherent since the Wiener increment of ${\bf c'}$ has no 
component parallel to ${\bf c'}$ itself.
Therefore, by decomposition along
$\bf n_2'$  = $\bf n_2'$ and $\bf n_3'$ we have

\begin{equation}
 \delta \mathcal{C}= 
\frac{\omega c}{\alpha}\left[   \,  
1 - \frac{{\bf v}\cdot{\bf c'}}{c^2}  \right] {\bf n'_2} \, dw_2
+
\frac{\omega c}{\alpha}\left[   \,  
(1-\alpha)\frac{({\bf v}\cdot {\bf n_3}')^2}{v^2}  
+  \left(1-\frac{{\bf v}\cdot{\bf c'}}{c^2} \right) 
\frac{{\bf c'} \cdot {\bf c}}{c^2} \right]  {\bf n_3'} \, dw_3 ,
\end{equation}
which given (\ref{cc'}) can be rewritten as
\begin{equation}
 \delta \mathcal{C}=  \frac{\omega \, c } {\alpha}  \, 
\left(1-\frac{{\bf v}\cdot{\bf c'} }{c^2} \right) \,
( {\bf n'_2} \, dw_2 + {\bf n'_3} \,dw_3). 
\label{bw}
\end{equation}

Finally, by $\delta {\bf c'}= \delta \mathcal{A}+\delta \mathcal{B}
+\delta \mathcal{C}$
we finally obtain

\begin{equation}
\delta {\bf c'}=  - \frac{\omega^2}{\alpha^2} 
\left[ 1-\frac{{\bf v} \cdot {\bf c'}}{c^2} \right]^2 {\bf c'}  dt
+ \frac{\omega \, c}{\alpha}  
\left( 1-\frac{{\bf v} \cdot {\bf c'}}{c^2} \right) 
({\bf n'_2}\, d w_2 + {\bf n'_3}\, dw_3),
\label{ito2}
\end{equation}
which is the equation (\ref{ito}) that we use in section 3 and which allows
to find out the equation (\ref{steq'}) which characterizes the general class 
of the light-speed processes.

\end{document}